\documentclass[aps,twocolumn,showpacs,footinbib]{revtex4}%
\usepackage[latin9]{inputenc}
\usepackage{amssymb}
\usepackage{amsmath}
\usepackage{amscd}
\usepackage{graphicx}
\usepackage{subfigure}
\usepackage{float}
\begin{document}
\draft
\title{Macroscopic superposition and entanglement for displaced thermal fields
induced by a single atom}
\author{Shi-Biao Zheng\thanks{%
E-mail: sbzheng@pub5.fz.fj.cn}}
\address{Department of Electronic Science and Applied Physics\\
Fuzhou University\\
Fuzhou 350002, P. R. China}
\date{\today }

\begin{abstract}
We show that a cavity field can evolve from an initial displaced mixed
thermal state to a macroscopic superpositions of displaced thermal states
via resonant interaction with a two-level atom. As a macroscopic system
(meter) is really in a mixed state before coupling with the microscopic
system at some temperature, our result is important for studying the quantum
measurement problem and decoherence under real conditions. For the two-mode
case, entanglement of displaced thermal states between the modes can be
obtained.
\end{abstract}


\vskip 0.5cm \maketitle \narrowtext

\section{INTRODUCTION}

The superposition principle is the most striking principle in quantum
mechanics. The macroscopic superposition states, known as Schr\"odinger cat
states [1], are of significance for exploring the boundary between quantum
and classical worlds and understanding the decoherence effect in quantum
information. The macroscopic superposition states also provide a test for
quantum measurement theory. In the measurement model, the coupling between a
macroscopic apparatus and a microscopic system results in their entanglement
and produces a quantum superposition state of the whole system.

The Jaynes-Cummings model (JCM), the simplest system in quantum optics
describing the interaction between light and matter, is an ideal system for
studying quantum interference effects [2,3]. It has been shown that if the
cavity field is initially in a coherent state with a large average photon
number superpositions of coherent states can be obtained in the resonant JCM
[4]. Recent advances in experiments involving the passage of single Rydberg
atoms through a superconducting cavity have turned the JCM from a
theoretical curiosity to a useful and testable enterprise [5]. Recently,
mesoscopic superpositions of two coherent states for a cavity field [6-8]
and the vibrational motion of a trapped ion [9] have been experimentally
demonstrated.

Previous researches on the macroscopic quantum-interference states
concentrate on superpositions of coherent states with different phases or
amplitudes since coherent states are considered to be the quantum states
close to classical ones. However, coherent states are not strictly classical
states since they are pure states. Before coupling with the microscopic
system, due to interaction with the environment a macroscopic apparatus is
in fact in a mixed state, instead of a pure state [10]. On the other hand,
thermal states, which do not exhibit any nonclassical property, are real
classical states. In fact, thermal states are true representation of the
state of a field at a finite temperature. The problem naturally arises: can
a thermal state evolve to a macroscopic superposition? Does such a mixed
state superposition exhibit quantum interference? How to detect the
coherence and decoherence of this state? These questions are closely related
with the quantum measurement process and decoherence under real conditions.

In this paper, we answer the above mentioned important questions, showing
that a cavity field can evolve to a macroscopic superposition through
interaction with a resonant two-level atom even if it is initially in a
thermal state. This is remarkable since it is believed that in order to
obtain a macroscopic superposition states one should initially prepare the
cavity mode in a pure coherent state. The coherence of the thermal state
superposition can be revealed by the collapses and induced revivals of the
Rabi oscillation. Our study opens a prospect for studying the quantum
measurement problem and exploring the quantum-classical transition at a
finite temperature. For the two-mode case, entanglement of thermal states
between two modes can be obtained. Our work is not only important for
studying how macroscopic quantum interference effects can persist at a
finite temperature, but also useful for quantum information processing with
mixed states.

The paper is organized as follows. In Sec.2, we show that an initial
displaced thermal state can evolve to superpositions of displaced thermal
states correlated with the atomic states via resonant atom-field
interaction. In Sec.3, we discuss the problem of detecting the coherence
between the displaced thermal states. In sec.4, we generalize the idea to
the two-mode case and show that entanglement of displaced thermal states
between the modes can be obtained. A summary appears in Sec.5.

\section{GENERATION OF SUPERPOSITIONS OF DISPLACED THERMAL STATES}

We consider a thermal field, whose state is described by the density
operator $\rho _{th}$%
\begin{equation}
\rho _{th}=\frac 1{\pi \stackrel{-}{n}_{th}}\int e^{-\left| \alpha \right|
^2/\stackrel{-}{n}_{th}}\left| \alpha \right\rangle \left\langle \alpha
\right| d^2\alpha ,
\end{equation}
where $\stackrel{-}{n}_{th}=1/(e^{\hbar \omega /k_BT}-1)$ is the mean
photon-number of the thermal field. We first displace the cavity field by an
amount $\alpha $, leading to the density operator $D(\alpha )\rho
_{th}D^{+}(\alpha )$, with $D(\alpha )$ being the displacement operator. We
here assume that $\alpha $ is a real number. We let a two-level atom
interact with the single-mode cavity field. In the rotating-wave
approximation, the Hamiltonian is (assuming $\hbar =1$)
\begin{equation}
H=g(a^{+}S^{-}+aS^{+}),
\end{equation}
where $S^{+}=\left| e\right\rangle \left\langle g\right| $, $S^{-}=\left|
g\right\rangle \left\langle e\right| ,$ with $\left| e\right\rangle $ and $%
\left| g\right\rangle $ being the excited and ground states of the atom, $%
a^{+}$ and $a$ are the creation and annihilation operators for the cavity
mode, and g is the atom-cavity coupling strength.

The evolution of the system is given by

\begin{equation}
\rho =U(t)D(\alpha )\left| \phi _a\right\rangle \left\langle \phi _a\right|
\otimes \rho _{th}D^{+}(\alpha )U^{+}(t),
\end{equation}
where
\begin{equation}
U(t)=e^{-iHt}.
\end{equation}
We can rewrite Eq. (3) as
\begin{equation}
\rho =D(\alpha )U_d(t)\left| \phi _a\right\rangle \left\langle \phi
_a\right| \otimes \rho _{th}U_d^{+}(t)D^{+}(\alpha ),
\end{equation}
where
\begin{equation}
U_d(t)=D^{+}(\alpha )U(t)D(\alpha ).
\end{equation}
Replacing Eq. (4) into (6) we obtain
\begin{equation}
U_d(t)=e^{-iH_dt},
\end{equation}
where
\begin{equation}
H_d=\frac 12g[(a^{+}+\alpha ^{*})S^{-}+(a+\alpha )S^{+}].
\end{equation}
Define the new atomic basis [11]

\begin{equation}
\left| +\right\rangle =\frac 1{\sqrt{2}}(\left| e\right\rangle +\left|
g\right\rangle ),\text{ }\left| -\right\rangle =\frac 1{\sqrt{2}}(\left|
e\right\rangle -\left| g\right\rangle ).
\end{equation}
Then we can rewrite $H_d$ as
\begin{equation}
H_d=\frac 12g[a^{+}(2\sigma _z+\sigma ^{+}-\sigma ^{-})+a(2\sigma _z+\sigma
^{-}-\sigma ^{+})]+2\Omega \sigma _z,
\end{equation}
where $\sigma _z=\frac 12(\left| +\right\rangle \left\langle +\right|
-\left| -\right\rangle \left\langle -\right| ),$ $\sigma ^{+}=\left|
+\right\rangle \left\langle -\right| $ , $\sigma ^{-}=\left| -\right\rangle
\left\langle +\right| ,$ and $\Omega =\alpha g$. We can rewrite the
displaced evolution operator $U_d(t)$ as
\begin{equation}
U_d(t)=e^{-i2\Omega \sigma _zt}e^{-iH_it},
\end{equation}
where
\begin{eqnarray}
H_i&=&\frac 12g[a^{+}(2\sigma _z+e^{i2\Omega t}\sigma
^{+}-e^{-i2\Omega t}\sigma ^{-})\cr&&+a(2\sigma _z+e^{-i2\Omega
t}\sigma ^{-}-e^{i2\Omega t}\sigma ^{+})]
\end{eqnarray}
Assuming that $\Omega \gg g$, i.e., $\alpha \gg 1$, we can neglect
the terms oscillating fast. Then $H_i$ reduces to
\begin{eqnarray}
H_i &=&g(a^{+}+a)\sigma _z  \nonumber \\
&=&\frac g2(a^{+}+a)(S^{-}+S^{+}).
\end{eqnarray}
Eq. (13) reveals the striking feature that, under the large displacement
condition, the dynamics of the normal JCM in the displaced picture is
described by the combination of the JCM and anti-JCM.

The displacement transformation alters the JCM evolution, resulting in the
competition of the excitation and deexcitation of the atomic state
accompanying the creation or annihilation of a photon in the initial thermal
field. After the displacement transformation, the excitation number $%
N_e=\left| e\right\rangle \left\langle e\right| +a^{+}a$ does not conserve
since the displacement operator, involving the competition between the
annihilation and creation operators, does not commute with the excitation
number operator.

We now assume that the atom is initially in the state $\left| g\right\rangle
.$ $\left| g\right\rangle $ can be rewritten as

\begin{equation}
\left| g\right\rangle =\frac 1{\sqrt{2}}(\left| +\right\rangle -\left|
-\right\rangle ).
\end{equation}
Using Eqs.(5) and (11), we obtain evolution of the system after an
interaction time $\tau $

\begin{equation}
\rho =D(\alpha )\left| \phi (\tau )\right\rangle \left\langle \phi (\tau
)\right| \otimes \rho _{th}D^{+}(\alpha ),
\end{equation}
\begin{eqnarray}
\left| \phi (\tau )\right\rangle &=&\frac 1{\sqrt{2}}[e^{-i\theta
}D\left( -\beta \right) \left| +\right\rangle +e^{i\theta }D\left(
\beta \right) \left| -\right\rangle \cr &=&\frac 12\{[e^{-i\theta
}D\left( -\beta \right) +e^{i\theta }D\left( \beta \right) ]\left|
e\right\rangle \cr&&+[e^{-i\theta }D\left( -\beta \right)
-e^{i\theta }D\left( \beta \right) ]\left| g\right\rangle \}.
\end{eqnarray}
where

\begin{eqnarray}
\theta &=&\Omega \tau , \\
\beta &=&ig\tau /2.  \nonumber
\end{eqnarray}
We can rewrite the density operator $\rho $ as
\begin{eqnarray}
\rho &=&\frac 14D(\alpha )[e^{-i\theta }D\left( -\beta \right)
+e^{i\theta }D\left( \beta \right) ]\rho _{th}[e^{-i\theta }D\left(
-\beta \right)\cr&& +e^{i\theta }D\left( \beta \right) ]D^{+}(\alpha
)\otimes \left| e\right\rangle \left\langle e\right|  +\frac
14D(\alpha )[e^{-i\theta }D\left( -\beta \right)\cr&& -e^{i\theta
}D\left( \beta \right) ]\rho _{th}[e^{i\theta }D(\beta )-e^{-i\theta
}D\left( -\beta \right) ]D^{+}(\alpha )\otimes \left| g\right\rangle
\left\langle g\right| \cr&& +\frac 14D(\alpha )[e^{-i\theta }D\left(
-\beta \right) +e^{i\theta }D\left( \beta \right) ]\rho
_{th}[e^{i\theta }D\left( \beta \right)\cr&& -e^{-i\theta }D\left(
-\beta \right) ]D^{+}(\alpha )\otimes \left| e\right\rangle
\left\langle g\right|  +\frac 14D(\alpha )[e^{-i\theta }D\left(
-\beta \right) \cr&&-e^{i\theta }D\left( \beta \right) ]\rho
_{th}[e^{-i\theta }D\left( -\beta \right) +e^{i\Omega \tau }D\left(
\beta \right) ]D^{+}(\alpha )\cr&&\otimes \left| g\right\rangle
\left\langle e\right| .
\end{eqnarray}
The unnormalized states $D(\alpha )[e^{-i\theta }D\left( -\beta \right) \pm
e^{i\theta }D\left( \beta \right) ]\rho _{th}[e^{i\theta }D\left( \beta
\right) \pm e^{-i\theta }D\left( -\beta \right) ]D^{+}(\alpha )$ are
superpositions of two displaced thermal states. The coherence arises from
the superposition of two different displacement operators. The average
photon-number of the displaced thermal state depends upon the amount of the
initial displacement: $\stackrel{-}{n}=$ $\stackrel{-}{n}_{th}+\left| \alpha
\right| ^2$. The quantum superposition persists in the classical limit $%
\left| \alpha \right| ^2\rightarrow \infty $.

\section{DETECTION OF THE COHERENCE BETWEEN THE DISPLACED THERMAL STATES}

In order to detect the coherence between two displaced thermal states we
could use the echo method proposed by Morigi et al. [12]. We divide the
whole duration into two parts. After an interaction time t, the evolution of
the system is given by the unitary operator $U_1=e^{-iHt}$. Then the atom
undergoes an instantaneous phase kick, corresponding to the application of
the inversion operator $S_z=(\left| e\right\rangle \left\langle e\right|
-\left| g\right\rangle \left\langle g\right| )$. For the remaining time $%
t^{^{\prime }}$, the JCM evolution operator is $U_1=e^{-iHt^{^{\prime }}}$.
The whole evolution operator is
\begin{equation}
U=U_2S_zU_1=S_ze^{-iH(t-t^{^{\prime }})}.
\end{equation}
We here have used the relation $S_zHS_z=-H$. Therefore, the phase kick leads
to the reversal of the unitary evolution of the system. After the duration
2t, the system evolves back to the initial state.

At the time $\tau $ ($\tau <t$), the probability for the atom in the state $%
\left| e\right\rangle $ is given by
\begin{eqnarray}
P_e &=&Tr\{D(\alpha )[e^{-i\theta }D\left( -\beta \right) \pm
e^{i\theta }D\left( \beta \right) ]\rho _{th}\cr&&[e^{i\theta
}D\left( \beta \right) \pm e^{-i\Omega \tau }D\left( -\beta \right)
]D^{+}(\alpha )\} \cr&=&\frac 12(1+e^{-(g\tau
)^2(\stackrel{-}{n}_{th}+2)/4}\cos (2\Omega \tau )].
\end{eqnarray}
The oscillation arises from the interference between the two displaced
thermal states $D(\alpha )D\left( -\beta \right) \rho _fD\left( \beta
\right) D\left( -\beta \right) D^{+}(\alpha )$ and $D(\alpha )D\left( \beta
\right) \rho _fD\left( -\beta \right) D^{+}(\alpha )$. The distance between
the two displaced states increases with the interaction time. When $g\tau
\sqrt{\stackrel{-}{n}_{th}+2}/2>1$, the two displaced states are
approximately orthogonal and thus the oscillation collapses. After the phase
kick, the evolution is reversed. At the time 2t, the two components merge
again into a single state and the atom returns to the initial state $\left|
g\right\rangle $. Then the Rabi oscillation resumes. The process can be
interpreted in term of complementarity. Due to the interaction with the
cavity mode, there exist two paths for the atom to reach a definite state,
with one path associated with the displaced thermal state $D(\alpha )D\left(
-\beta \right) \rho _fD\left( \beta \right) D^{+}(\alpha )$ and the other
associated with $D(\alpha )D\left( \beta \right) \rho _fD\left( -\beta
\right) D^{+}(\alpha )$. In the case $g\tau /2\ll 1$, the difference between
the two displaced thermal states is negligible and thus no path information
can be obtained. With the increase of the interaction time, the two
displaced thermal states become distinguishable and the path information is
recorded on the cavity mode, destructing the interference. After the phase
kick, the distance between the two displaced thermal states decreases. At
time 2t, the two components are recombined and the path information is
erased, resulting the reappearance of the interference. The contrast of the
oscillation is independent of the value of $\alpha $. Therefore, collapses
and revivals occur for the displaced thermal states even when $\stackrel{-}{n%
}\rightarrow \infty $.

With the decoherence being considered, the contrast of the oscillation is
reduced. The initial displaced thermal state can be expressed in terms of
the Fock states
\begin{equation}
D(\alpha )\rho _{th}D^{+}(\alpha )=\sum_{m,n=0}^\infty \rho _{m,n}\left|
m\right\rangle \left\langle n\right| ,
\end{equation}
where
\begin{equation}
\rho _{m,n}=\sum_{\nu =0}^\infty \frac{\stackrel{-}{n}^\nu }{(1+\stackrel{-}{%
n})^{\nu +1}}C_{\nu ,m}C_{\nu ,n}^{*},
\end{equation}
\begin{eqnarray}
C_{\nu ,m}&=&\sqrt{\nu !m!}e^{-\alpha ^2/2}\sum_{\nu =0}^\infty
\frac 1{l!(\nu -l)!(m-l)!}\cr&&(-\alpha )^{\nu -l}\alpha
^{m-l},l\leq m
\end{eqnarray}
Set $\kappa \ll g$ and $\kappa t\ll 1$, with $\kappa $ being the
cavity decay rate. At the time T, due to the cavity decay the
contrast of the oscillation is reduced by

\begin{eqnarray}
&&\sum_{n=0}^\infty \kappa \sqrt{1+2\stackrel{-}{n}_{th}}\rho _{nn}\{\frac t4%
(2n-1)+\frac{\sin (gt\sqrt{n})}{4g\sqrt{n}}\cr&&-\frac{\sin (gt\sqrt{n-1})}{4g%
\sqrt{n-1}}
-\frac 1{4g}[\sqrt{n}(4n-3)\sin (gt\sqrt{n})\cr&&\cos (gt\sqrt{n-1})-\sqrt{n-1}%
(4n-1)\sin (gt\sqrt{n-1})\cr&&\cos (gt\sqrt{n})]\}.
\end{eqnarray}

\section{GENERATION OF ENTANGLEMENT BETWEEN TWO THERMAL CAVITY MODES}

We note the idea can also be generalized to generate entanglement between
two thermal cavity modes. We consider two degenerate cavity modes initially
in the thermal states with the density operators $\rho _{th,1}$ and $\rho
_{th,2}$. We first displace each cavity mode by a large amount $\alpha $ ($%
\alpha \gg 1)$, leading to the density operator $D_1(\alpha )\rho
_{th,1}D_1^{+}(\alpha )\otimes D_2(\alpha )\rho _{th,2}D_2^{+}(\alpha )$.
The two-level atom resonantly interact with the two modes. In the
interaction picture, the Hamiltonian is
\begin{equation}
H=(g_1a^{+}+g_2b^{+})S^{-}+(g_1a+g_2b)S^{+}],
\end{equation}
where $a$ and $b$ are the annihilation operators for the two modes, and $g_1$
and $g_2$ are the corresponding coupling strengths. After the atom exits the
cavity the whole system is in the state
\begin{equation}
\rho =D(\alpha ,\alpha )\left| \phi (\tau )\right\rangle \left\langle \phi
(\tau )\right| \otimes \rho _{th,1}\rho _{th,2}D^{+}(\alpha ,\alpha ),
\end{equation}
where
\begin{equation}
\left| \phi (\tau )\right\rangle =\frac 1{\sqrt{2}}[e^{-2i\theta }D(-\beta
_1,-\beta _2)\left| +\right\rangle +e^{2i\theta }D(\beta _1,\beta _2)\left|
-\right\rangle ,
\end{equation}
\begin{equation}
D(\beta _1,\beta _2)=D_1(\beta _1)D_2(\beta _2),
\end{equation}
\begin{eqnarray}
\beta _1 &=&ig_1\tau /2, \\
\beta _2 &=&ig_2\tau /2.  \nonumber
\end{eqnarray}
We can rewrite the state as
\begin{eqnarray}
\rho &=&\frac 14D(\alpha ,\alpha )D_{+}\rho _{th,1}\otimes \rho
_{th,2}\ D_{+}D^{+}(\alpha ,\alpha )\otimes \left| e\right\rangle
\left\langle e\right| \cr&& -\frac 14D(\alpha ,\alpha )D_{-}\rho
_{th,1}\otimes \rho _{th,2}\ D_{-}D^{+}(\alpha ,\alpha )\otimes
\left| g\right\rangle \left\langle g\right|  \nonumber \cr && -\frac
14D(\alpha ,\alpha )D_{+}\rho _{th,1}\otimes \rho _{th,2}\
D_{-}D^{+}(\alpha ,\alpha )\otimes \left| e\right\rangle
\left\langle g\right|  \nonumber \cr &&+\frac 14D(\alpha ,\alpha
)D_{-}\rho _{th,1}\otimes \rho _{th,2}\ D_{+}D^{+}(\alpha ,\alpha
)\otimes \left| g\right\rangle \left\langle e\right| ,
\end{eqnarray}
where
\begin{eqnarray}
D_{+} &=&e^{-2i\theta }D(-\beta _1,-\beta _2)+e^{2i\theta }D(\beta _1,\beta
_2) \\
D_{-} &=&e^{-2i\theta }D(-\beta _1,-\beta _2)-e^{2i\theta }D(\beta _1,\beta
_2)  \nonumber
\end{eqnarray}
The states $D(\alpha ,\alpha )D_{\pm }\rho _{th,1}\otimes \rho _{th,2}\
D_{\pm }D^{+}(\alpha ,\alpha )$ are entangled displaced thermal states. The
entanglement results from the superposition of the two-mode displacement
operators $D(-\beta _1,-\beta _2)$ and $D(\beta _1,\beta _2)$. At the time $%
\tau $, the probability for the atom in the state $\left| e\right\rangle $
is given by
\begin{equation}
P_e=\frac 12(1+e^{-[(g_1\tau )^2+(g_2\tau )^2](\stackrel{-}{n}+2)/4}\cos
(4\Omega \tau )].
\end{equation}
The oscillation arises from the interference between the two two-mode
thermal states $D(\alpha ,\alpha )D(-\beta _1,-\beta _2)\rho _{th,1}\otimes
\rho _{th,2}\ D(\beta _1,\beta _2)D^{+}(\alpha ,\alpha )$ and $D(\alpha
,\alpha )D(\beta _1,\beta _2)\rho _{th,1}\otimes \rho _{th,2}\ D(-\beta
_1,-\beta _2)D^{+}(\alpha ,\alpha )$. The distance between the two two-mode
displacement operators does not decrease as the average photon-numbers of
the displaced thermal states increases. Therefore, the entanglement survives
when the average photon-numbers of the two fields approach infinity. The
idea opens promising prospects for entangling two macroscopic mixed systems
and investigate the decoherence. The idea can be easily generalized to
produce entanglement for two or more thermal fields located in separated
cavities.

\section{SUMMARY}

In conclusion, we show that the macroscopic thermal state superposition can
be induced by a resonant two-level atom for the first time. The quantum
coherence can be revealed by the collapses and induced revivals of the Rabi
oscillation. In contrast with previous studies concerntrating on pure
coherent states, our research provides a way for investigating the quantum
measurement model and decoherence phenomena under real conditions. For the
two mode case, the quantum entanglement between two thermal cavity modes can
be obtained. The entanglement survives even if the average photon-numbers of
the modes go to infinity. The required experimental techniques for
demonstrating the idea including the passage of single Rydberg atoms through
a high-quality cavity, displacement, phase kick, and atomic state
measurement. All these techniques are presently available [6-8] and thus the
implementation of the idea appears experimentally feasible.

This work was supported by the National Fundamental Research Program Under
Grant No. 2001CB309300, and the National Natural Science Foundation of China
under Grant No. 10674025.

\end{document}